\documentclass[12pt]{emulateapj}
\usepackage[dvips]{color} 
\usepackage{lscape}
\usepackage{amssymb}
\newcommand{\Msun}{{{\rm M}_\odot}}

\newcommand{\jj}{{\rm j}}

\newcommand{\ahiso}{$\alpha_{h, {\rm iso}}$}
\newcommand{\aliso}{$\alpha_{l, {\rm iso}}$}
\def\simless{\mathbin{\lower 3pt\hbox
{$\rlap{\raise 5pt\hbox{$\char'074$}}\mathchar"7218$}}}   %< or of order
\def\simmore{\mathbin{\lower 3pt\hbox
{$\rlap{\raise 5pt\hbox{$\char'076$}}\mathchar"7218$}}}   %> or of order

\slugcomment{accepted for publication in ApJ}

\shorttitle{Relativistic Jet Eruption from Collapsars}
\shortauthors{Mizuta \& Aloy}

\begin{document}

\title{Angular Energy Distribution of Collapsar-Jets}

\author{Akira Mizuta\altaffilmark{1}, 
and 
Miguel A. Aloy\altaffilmark{2},
}

\altaffiltext{1}{Center for Frontier Science, Chiba University
  Yayoi-cho 1-33, Inage-ku, Chiba, 263-8522 Japan}
\altaffiltext{2}{Departamento de Astronom\'ia y Astrofis\'ica,
  Universidad de Valencia, 46100-Burjassot (Valencia), Spain}
\email{AM, e-mail: mizuta@cfs.chiba-u.ac.jp}

\begin{abstract}
Collapsars are fast-spinning, massive stars, whose core collapse
liberates an energy, that can be channeled in the form of
ultrarelativistic jets. These jets transport the energy from the
collapsed core to large distances, where it is dissipated in the
form of long-duration gamma-ray bursts. In this paper we study the
dynamics of ultrarelativistic jets produced in collapsars. Also we
extrapolate our results to infer the angular energy distribution of
the produced outflows in the afterglow phase. Our main focus is to
look for global energetical properties which can be imprinted by the
different structure of different progenitor stars. Thus, we employ a
number of pre-supernova, stellar models (with distinct masses and
metallicities), and inject in all of them jets with fixed initial
conditions. We assume that at the injection nozzle, the jet is
mildly relativistic (Lorentz factor $\sim 5$), has a finite
half-opening angle ($5^\circ$), and carries a power of
$10^{51}\,$erg\,s$^{-1}$. In all cases, well collimated jets
propagate through the progenitor, blowing a high pressure and high
temperature cocoon. These jets arrive intact to the stellar surface
and break out of it.  A large Lorentz factor region $\Gamma\simmore
100$ develops well before the jet reaches the surface of the star,
in the unshocked part of the beam, located between the injection
nozzle and the first recollimation shock. These high values of
$\Gamma$ are possible because the finite opening angle of the jet
allows for free expansion towards the radial direction.  We find a
strong correlation between the angular energy distribution of the
jet, after its eruption from the progenitor surface, and the
mass of the progenitors.  The angular energy distribution of the
jets from light progenitor models is steeper than that of the jets
injected in more massive progenitor stars. This trend is also
imprinted in the angular distribution of isotropic equivalent
energy.
\end{abstract}


%%%%%%%%%%%%%%%%%%%%%%%%%%%%%%%%%%%%%%%%%%%%%%%%%%%%%%%%%%%%%%%%%%%%%%
%
\section{INTRODUCTION}
%
%%%%%%%%%%%%%%%%%%%%%%%%%%%%%%%%%%%%%%%%%%%%%%%%%%%%%%%%%%%%%%%%%%%%%%
Recent observations of gamma-ray bursts (GRBs) suggest that long
duration GRBs and type Ib/c supernova (SN) explosions are tightly
connected.  For example, SN1998bw was observed in the positional error
box of GRB980425 \citep{Galama98}.  In this case the GRB/SN
association was based on the spatial and temporal coincidence of both
events.  The most remarkable example of long GRB/SN link came in 2003,
when the spectra of both the GRB030329 afterglow and of the SN2003dh
were measured, since the burst happened closeby and it was quite
bright.  The supernova spectrum, which includes many complex lines,
gradually appeared from the decaying afterglow spectrum after a few
tens of days from the burst.  The spectrum of SN2003dh after about a
month from the explosion is quite similar with that of SN1998bw at the
same stage.  Both SN1998bw and SN2003dh are type Ic supernovae
\citep{Iwamoto98,Stanek03} whose progenitor had lost the hydrogen and
the helium envelopes during the pre-supernova stage.  They are also
categorized within a special class of supernova explosions, so-called
hypernovae, whose explosion energy is about ten times higher, i.e.,
$\sim 10^{52}\ \mbox{erg}$, than that of ordinary supernova.
Indeed, our common view is that most long-lasting GRBs are produced by
core-collapse supernovae akin to SN1998bw.

GRBs are not exclusively linked to hypernovae.
For instance, the long lasting burst ($t_d\sim 2\times 10^3$ s)
GRB060218 (or XRF060218)
was associated with the type Ic \object{SN2006aj}, which is not a
hypernova.  The explosion energy falls within the regular range for
type Ic events \citep{Campana06}.  \cite{Mazzali06} argued that the
progenitor star may not form a black hole but a neutron star, since
the estimated mass of the progenitor during the main sequence is $\sim
20 M_\odot$.  On the other hand, there are recent examples of cases of
long-lasting GRBs (\object{GRB060505} and \object{GRB060614}) where no
supernova signature was observed at all even if, considering the
distance, the observational trace of a supernova should have been
detected \citep{Fynbo06,DellaValle06,GalYam06}.

Though a variety of long duration GRBs which have a strong connection
with SNs are observed, we still lack of a complete picture of the
processes by which a sizable fraction of the energy involved in a type
Ib/c SN explosion is tapped in a relatively narrow channel, and
produces a GRB at a large distance from the original site of
generation.  \citet{Rees92} proposed that the death of massive stars
can be an origin of GRBs.  \cite{Woosley93} introduced the collapsar
model to account for the progenitor system of long GRBs.  According to
this model, a non-spherical outflow could be formed from the deep
inside of the progenitor where a black hole or a proto-neutron star is
born as a result of the collapse of the iron core \citep{MacFadyen99}.
If the specific angular momentum of the iron core is sufficiently
large, an accretion torus may develop around the central compact
object.  The system formed by a central object and an accretion disk
has the potential to launch a bipolar outflow.

  The mechanisms proposed to extract energy out of such
central engines are basically two \citep[e.g.,][]{AO07}: thermal or
hydromagnetic.  Thermal mechanisms rely on depositing a considerable
amount of thermal energy in the vicinity of the rotation axis of the
system, just above the poles of the central compact object, where a
low density funnel has develop in the course of the evolution. The
accretion energy of the hot torus is converted into a copious flux of
neutrinos $\nu$ and anti-neutrinos ${\bar \nu}$. From the $\nu{\bar
  \nu}$-annihilation a hot $e^+e^-$-plasma results. In its turn,
$e^+e^-$-pairs annihilate yielding a {\em fireball} of high energy
photons. The conversion of the thermal energy of the fireball into
kinetic energy partly determines the subsequent evolution of the
plasma.  It accelerates to ultrarelativistic speeds (reaching Lorentz
factors of $\sim 50$ \citealt{Aloy00}) while, at the same time,
interacts with the progenitor system. Alternatively, MHD process may
tap a fraction of the rotational energy of the BH or of the accretion
disk to form an outflow
\citep{Proga03,Mizuno04,McKinney06,Nagataki07,Komissarov07,Takiwaki07}.

A number of works have dealt with the hydrodynamic properties of the
outflows generated in collapsar progenitors as they propagate through
the progenitor system (and in some cases beyond the surface of the
progenitor star).  The problem is addressed by means of numerical
relativistic hydrodynamic simulations with different degrees of
complexity (\citealt{Zhang03,Zhang04,Mizuta06,Morsony07,Tominaga07}),
which assume that a quasi-steady momentum flux has been produced at a
certain distance from the region where the energy is released
(independent of which is the actual energy extraction mechanism -MHD
or thermal-). Therefore, such numerical works assume the existence of
a nozzle through which the injection of a supersonic jet is produced,
and put their focus on the modification of the morphology and of the
dynamics of collimated outflows as they travel through the progenitor
star. Depending on the exact inflow conditions, a variety of different
outflows result. For instance, \citet{Mizuta06} finds a whole spectrum
of outflows ranging from collimated, relativistic jets to poorly
collimated expanding winds. Thus, \citeauthor{Mizuta06} argue
that such a variety of resulting outflows supports the idea that the
same collapsar scenario can yield a number of different phenomena (in
agreement with the previous ideas of unification of high energy
transients, e.g., \citealp{Ramirez-Ruiz02b}), such as, GRBs, X-ray
rich GRBs, X-ray Flashes, and normal supernovae.

In the prompt GRB phase we observe the emission of high energy photons
from an ultrarelativistic outflow, which is generated at a distance
$10^{13-15}$\,cm, very far from the central engine.  Due to the large
optical thickness of the outflow at scales comparable to that of the
progenitor, we have not observed any electromagnetic emission directly
from the progenitor so far, except, perhaps, some precursor activity
\citep{Woosley00,Ramirez-Ruiz02}.  Since the direct detection of
progenitors of GRBs is nowadays impossible (if they happen at
cosmological distances), the observation of the association between
GRBs and SNe probably provides the best clue to understand the
progenitors of the GRBs.  Though more than 100 GRBs per year are
identified, most of them occur far away.  Thus, it is technically
impossible to identify their accompanying supernovae.  Another clue
regarding the nature of the GRB progenitors comes from the environment
and the host galaxies in which the burst takes place. Both
  clues indicate that long-duration GRBs are associated with the
death of the most massive stars \citep{Fruchter06}.  The typical hosts
of long GRBs are star-forming, low metallicity galaxies (with an star
formation rate $\sim 10^3\,M_\odot$y$^{-1}$;
\citealt{Berger01,Frail02}) but bluer than typical starburst galaxies,
with little dust \citep{LeFloch04}, and lower masses than current
ellipticals, i.e., they correspond to the typical
environments of formation of massive stars.  \cite{Fruchter06} also
conclude, that the host galaxies of the GRBs are significantly fainter
and more irregular than the hosts of core-collapse supernovae.

A very interesting question is weather it is possible to get any
information on the nature of a GRB progenitor from the observation of
the afterglow emission.  One possibility is to look at the angular
distribution of integrated energy per unit of solid angle, as observed
in the afterglow phase of the burst.  \citet{Lazzati05} estimated
theoretically such an angular distribution assuming, that the kinetic
energy of the jet is converted to thermal energy in the cocoon,
till the head of the jet reaches the progenitor surface.  The cocoon
originates from jet material which crosses through the terminal strong
shock of the collimated outflow and moves away from the center of the
progenitor surrounding the beam of the jet.  A fat cocoon develops for
light jets, i.e., jets whose rest-mass density is much lower than that
of the ambient gas into which the jet propagates.  When the jet breaks
out the progenitor surface, the thermal energy is released to a
  low-density inter stellar medium (ISM).  \cite{Lazzati05}
concluded, that the energy distribution per solid angle
  ($dE/d\Omega$) of the jet displays a $\theta^{-2}$ dependence with
  the viewing angle after its eruption through the progenitor
surface.  Recently, \citet{Morsony07} have used hydrodynamic
simulations to test the theoretical prediction of
\citeauthor{Lazzati05}, and found that their numerical models do not
follow the inferred theoretical angular energy distribution.

In this paper we also try to verify the analytic relation for the
angular dependence of the energy with the polar angle that was
proposed by \cite{Lazzati05}. We explore a parameter space different
from that of \cite{Morsony07} in order to compute the dependence of
the angular energy distribution on the structure of the
progenitor. The progenitor models are built upon the
pre-supernova models of \cite{Woosley06}.  Along the way, we also
characterize the hydrodynamic properties of relativistic jets
propagating through different progenitor stars.

The paper is organized as follows. We describe our physical model, the
choice of stellar progenitors, and relevant numerical details in
Sect.~\ref{model}. In the appendix, we provide a study of a
  selected sample of models in order to justify our choice of
  numerical resolution and the effects it has on our conclussions.
The dynamics of the injected bipolar outflows, and the extrapolated
angular energy distribution in the afterglow phase is considered in
Sec.~\ref{results}. Finally, we discuss our results and write down the
conclusions of this work in Sect.~\ref{sec:conclusions}.

%%%%%%%%%%%%%%%%%%%%%%%%%%%%%%%%%%%%%%%%%%%%%%%%%%%%%%%%%%%%%%%%%%%%%%
%
\section{MODEL}
\label{model}
%
%%%%%%%%%%%%%%%%%%%%%%%%%%%%%%%%%%%%%%%%%%%%%%%%%%%%%%%%%%%%%%%%%%%%%%

We will investigate the dependence of the properties of
relativistic jets, injected in a pre-supernova stellar model at a
certain distance from the center, using relativistic hydrodynamic
simulations. In Sect.~\ref{sec:progenitor} we show the different
stellar progenitors used in this study. We provide the technical
details of the numerical simulations in
Sect.~\ref{sec:numerics}. Finally, in Sect.~\ref{sec:jet} we specify
the physical conditions used to inject relativistic jets in the
pre-supernova progenitors described above.

%%%%%%%%%%%%%%%%%%%%%%%%%%%%%%%%%%%%%%%%%%%%%%%%%%%%%%%%%%%%%%%%%%%%%%
%
\subsection{Progenitors}
\label{sec:progenitor}
%
%%%%%%%%%%%%%%%%%%%%%%%%%%%%%%%%%%%%%%%%%%%%%%%%%%%%%%%%%%%%%%%%%%%%%%

In the last years, some detailed calculations of stellar evolution of
massive stars have been done including the effects of initial angular
momentum, dynamo, metallicity, and mass loss rate
\citep{Yoon05,Yoon06,Woosley06}. According to these studies,
the metallicity of the progenitor strongly affects the evolution of
the angular momentum distribution at the pre-supernova stage, in
such a way, that low metallicity is preferred to obtain a large
angular momentum in the core of the progenitor.

For the purposes of this work, we employ some of the pre-supernova
models computed by \citet{Woosley06}.  We stick to the same naming
convention than the former authors, and consider several sets of
models (Tab.~\ref{tab:progenitor}).  The first group corresponds to
the HE16-series of 16 models of \citeauthor{Woosley06}, which include
progenitor stars for which $16 M_\odot$ bare helium cores are evolved,
that have solar metallicity, and different amounts of initial angular
momentum, dynamo effects, mass-loss rates, etc.  The last three models
of Tab.~\ref{tab:progenitor}, 16OC, 16TB, and 16TC form the second
group of progenitors.  They correspond to stars with the same initial
mass as those of the first group but with a smaller metallicity
($Z_{\rm 16OC}=0.1Z_\odot$, $Z_{\rm 16TB}=Z_{\rm 16TC}=0.01Z_\odot$).
The second set of low-metallicity models has been chosen
among the many other possibilities available because their radius, at
the pre-supernova stage, are the smallest among all other
low-metallicity progenitors (in all cases, their stellar radii
are $R_* < 10^{11}$\,cm).

\begin{figure}
\begin{center}
\includegraphics[scale=0.35]{{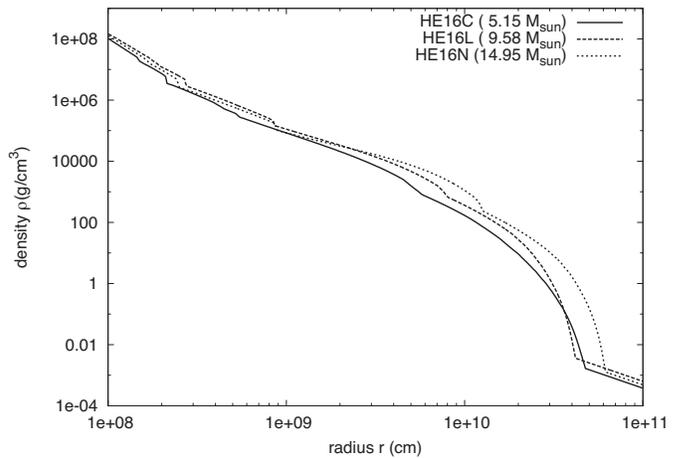}}
\caption{Radial mass profiles of models HE16C, HE16L and HE16N.}
\label{fig:1dradialmass}
\end{center}
\end{figure}

\begin{figure}
\begin{center}
\includegraphics[scale=0.35]{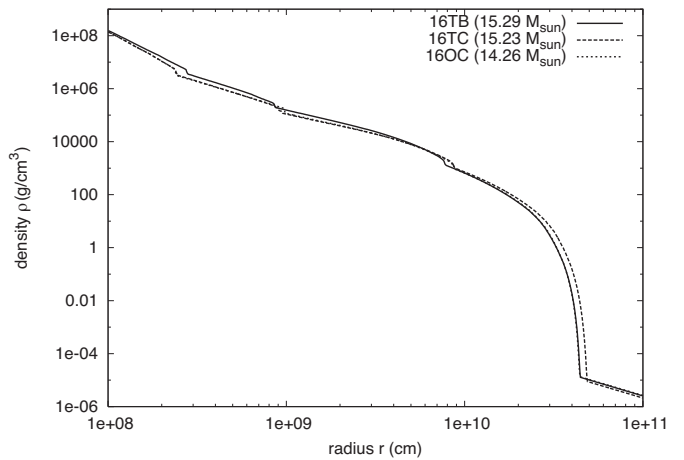}
\caption{Radial mass profiles of models 16TB, 16TC and 16OC.}
\label{fig:1dradialmass2}
\end{center}
\end{figure}

Figure~\ref{fig:1dradialmass} shows the radial mass profiles of the
models HE16C, HE16L, and HE16N. HE16C is representative of progenitors
whose pre-supernova mass is small $M_{\rm HE16C}\simeq 5 \Msun$ (due
to the vigorous mass loss rate in the late phases of its evolution;
Tab.~\ref{tab:progenitor}).  Other members of this group of low
pre-supernova mass are HE16B, HE16J, and HE16K. We will refer to this
group as type-L.  The model HE16N belongs to the group of more massive
progenitors ($M_{\rm HE16C}\simeq 15 \Msun$), to which we will refer
as type-H models.  Finally, the model HE16L (also HE16D) falls in the
middle of these two groups (members of this group will be called
type-M models).  Its total mass is about $9.5\Msun$.  Figure
\ref{fig:1dradialmass2} shows radial mass density profiles of the low
metallicity models 16TB, 16TC, and 16OC.  Though the mass of the
models is similar ($\sim 15.3\Msun$), the density profiles are
slightly different.  In total, 19 models are considered in this study
(Tab.~\ref{tab:progenitor}).  We neglect for the progenitors any
deviation from spherical symmetry arising from the rotation of the
models.  Therefore, for the study we present here, each progenitor
differs from the other mainly in its total mass, radius, and mass
density profile at the pre-supernova stage.

Though a nonspherical structure is expected around the black hole due
to the rapid rotation of the progenitor, it is reasonable to assume
spherical symmetry for the envelopes of the progenitor for radial
distances $r \gtrsim 10^8$\,cm, which is where we put the innermost
radial boundary in the numerical simulations of this study.  Thus, we
only take from the models of \citet{Woosley06} the radial density and
the radial velocity profiles which result by the end of the
pre-supernova evolution.  We assume that the pressure of the
progenitor is very low, or, equivalently, that the initial
specific internal energy ($\epsilon$) is set to be very low
($\epsilon/c^2=10^{-6}$, where $c$ is the speed of light).

Both the gravitational force produced by the central compact
object, and the progenitor self gravity are ignored, since
the timescale for the outflows to cross the progenitor and to break
out from the surface of the progenitor, $\sim 3$s, is much shorter
than free fall timescale for the stellar envelopes.

We extend the radial mass density profile to the outside of the
progenitor up to the outer computational boundary located at $r_{\rm
  max}=10^{11}$\,cm.  For the models which barely loose mass
during the latest stages of their evolution, the rest-mass density is
assumed to be uniform $(\rho_{\rm ISM}=10^{-6}\,\mbox{g cm}^{-3})$ and
much smaller than that at the progenitor surface.  If the progenitor
star has a non-null mass loss rate (i.e., if the parameter
  $a>0.01$, according to the nomenclature of \citealp{Woosley06}), we
take a $r^{-2}$ dependence in radial mass-density profiles from the
surface of the progenitor (see Fig.~\ref{fig:1dradialmass}).

%%%%%%%%%%%%%%%%%%%%%%%%%%%%%%%%%%%%%%%%%%%%%%%%%%%%%%%%%%%%%%%%%%%%%%
%
\subsection{Computational Domain and Basic Equations}
\label{sec:numerics}
%
%%%%%%%%%%%%%%%%%%%%%%%%%%%%%%%%%%%%%%%%%%%%%%%%%%%%%%%%%%%%%%%%%%%%%%

We map the spherically symmetric progenitor models of
Sect.~\ref{sec:numerics} into a two-dimensional grid in spherical
coordinates ($r\times \theta$).  We assume that our models are axial
and equatorially symmetric and, therefore, specify reflection boundary
conditions at the polar axis ($\theta=0^\circ$) and at the equator
($\theta=90^\circ$).  The radial grid consists of $N_r=1000$ points,
uniformly spaced in $\log{r}$, which extends from $r_{\rm
  min}=10^8$\,cm to $r_{\rm max}=10^{11}$\,cm.  The smallest radial
grid spacing, besides $r_{\rm min}$, is $\Delta r_{\rm min}=10^6$\,cm,
while the largest one, besides $r_{\rm max}$, is $\Delta r_{\rm
  max}=6.5\times 10^8$\,cm.  The resolution we have choosen
  here represents a trade-off between accuracy and feasibility of the
  numerical simulations, as we discuss in the Appendix.  Free outflow
(i.e., zero gradient) boundary conditions are set at $r=r_{\rm min}$
and $r_{\rm max}$.  The polar grid has $N_\theta=180$ grid points
uniformly spaced in the range $0^{\circ}<\theta <90^{\circ}$, ($\Delta
\theta =0.5^{\circ}$).  We use the same 2D special relativistic
hydrodynamic code of \cite{Mizuta04,Mizuta06} to perform our
simulations.  The code provides 3rd order accuracy in both space and
time, by applying a PPM intra-cell interpolation and a TVD-Runge Kutta
time integration.  For the sake of simplicity, we employ an ideal gas
equation of state ($p=(\gamma-1)\rho\epsilon$) with uniform
adiabatic index $\gamma=4/3$, where $p$, $\rho$ and $\epsilon$ are the
pressure and the rest-mass density, respectively.

%%%%%%%%%%%%%%%%%%%%%%%%%%%%%%%%%%%%%%%%%%%%%%%%%%%%%%%%%%%%%%%%%%%%%%
%
\subsection{Jet Injection Conditions}
\label{sec:jet}
%
%%%%%%%%%%%%%%%%%%%%%%%%%%%%%%%%%%%%%%%%%%%%%%%%%%%%%%%%%%%%%%%%%%%%%%

We assume that a jet has been generated by the central engine, and
that at a certain distance, quasi-steady injection conditions are
settled through a well defined circular nozzle. Thus, we inject
plasma, in the radial direction, through the innermost radial boundary
at $r=r_{\min}$ in a cone of half-opening angle
$\theta_\jj=5^\circ$. The jet injection proceeds for a period $t_{\rm
  inj}=4\,$s. We parametrize the outflowing plasma by assuming that it
is hot (we set $\epsilon_\jj/c^2=30$) and moderately relativistic (the
Lorentz factor being $\Gamma_{\jj,0}=5$). We adopt the convention that
the parameters of the outflow at the injection point are named with a
subscript 'j'. Because of the conversion of thermal-to-kinetic energy,
the injected flows have the potential to accelerate to bulk Lorentz
factors larger than 100 \citep{Mizuta06}.  During the first 3\,s, the
power of the injected outflow is $L_{\jj,0} \equiv
\rho_\jj\Gamma_{\jj,0}{v_r}_\jj (h_\jj \Gamma_{\jj,0}-1)c^2\Delta
S=10^{51}\mbox{ erg s}^{-1}$, where $\Delta S$ is the area of the
injection surface, $h(\equiv 1+\epsilon/c^2 + p/\rho c^2)$ is the
specific enthalpy, and $v_r$ is the radial component of the
3-velocity.  The density and pressure of the injected outflow are
obtained by setting $\Gamma_{\jj,0}$, $\epsilon_\jj$, $L_{\jj,0}$,
$\theta_{\jj}$, and $r_{\rm min}$. We fix $L_{\jj,0}=10^{51}\mbox{ erg
  s}^{-1}$, which is higher than that adopted in previous studies
\citep{Zhang03,Zhang04,Morsony07}.  The total injected energy is
several times $10^{51}$\,erg.  Since the main purpose of this study is
to see the jet propagation and expansion of the cocoon into the
interstellar medium after the shock breakout, we adopt this power to
obtain a rapid propagation of the jet in the progenitor.  This fast
propagation is necessary to be consistent with the fact that we
neglect the self-gravity of the star.  If the jet crosses the
progenitor much faster than the typical hydrodynamic timescale in the
system, the progenitor remains roughly unchanged during the complete
jet propagation through it and, therefore, we do not need to care
about the progenitor evolution during such short timescales.

After the initial phase of constant kinetic power injection, both the
kinetic power and the injection Lorentz factor are linearly decreased
according to the laws $L_\jj(t) = \max \{L_{\jj,0}(4-t),
10^{49}\,\mbox{erg\,s}^{-1}\}$, and $\Gamma_\jj(t) =
\max \{\Gamma_{\jj,0}+12-4t, 1.01\}$, respectively, for
$3\,\mbox{s}<t<4\,\mbox{s}$. In this period of decaying injection
power, the specific energy is kept fixed to the same value as it had
at $t=0$, and the density and the pressure are obtained from the other
parameters (as in the constant injection power phase).  After
$t=4$\,s, the flow injection ceases.

With the parametrization considered above, the rest-mass density
($\rho_{\jj,0}$) during the constant power phase is 154\,g\,cm$^{-3}$.
Since the rest-mass density of the progenitor around the inner computational
boundary is $\sim10^8$\,g\,cm$^{-3}$,
the injected outflow is initially much lighter
than medium in which it is injected.
Thus it is expected that the jet propagation velocity across
the progenitor is smaller than the speed of light,
and one naturally expects to generate relatively
 thick cocoons surrounding the beam of the jet.

%%%%%%%%%%%%%%%%%%%%%%%%%%%%%%%%%%%%%%%%%%%%%%%%%%%%%%%%%%%%%%%%%%%%%%
%
\section{RESULTS}
\label{results}
%
%%%%%%%%%%%%%%%%%%%%%%%%%%%%%%%%%%%%%%%%%%%%%%%%%%%%%%%%%%%%%%%%%%%%%%

%%%%%%%%%%%%%%%%%%%%%%%%%%%%%%%%%%%%%%%%%%%%%%%%%%%%%%%%%%%%%%%%%%%%%%
%
\subsection{Dynamics}
\label{sec:dynamics}
%
%%%%%%%%%%%%%%%%%%%%%%%%%%%%%%%%%%%%%%%%%%%%%%%%%%%%%%%%%%%%%%%%%%%%%%

The dynamical evolution of our jet models can be split in two phases.
The first one happens during the period in which the jet drills its
way through the progenitor star.  The second one shows up latter, when
the jet breaks out of the stellar surface.  The dynamics of our models
during this two phases if roughly similar to that outlined by some
previous works (e.g.,
\citealp{Aloy00,Zhang03,Zhang04,Mizuta06,Morsony07}) and, therefore,
we limit ourselves here to provide a shallow description of the most
salient features.

\begin{figure}
\begin{center}
\includegraphics[scale=0.5]{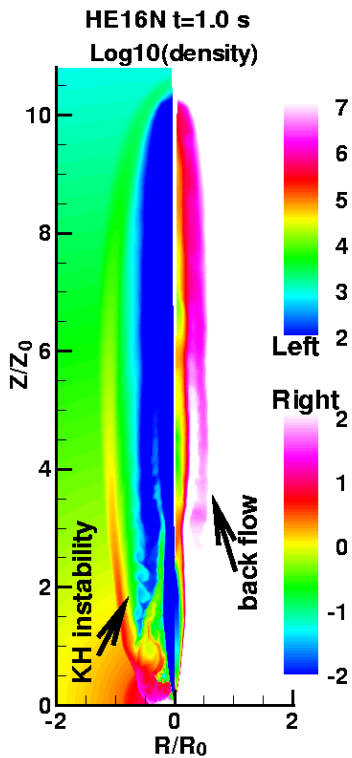}
\caption{Snapshot (at $t=1.0$\,s) of the rest-mass density of
    model HE16N. The left and right panels are shown with different
    color scales in order to outline the most salient features of the
    cocoon and the cavity drilled by the jet (left), and to show more
    clearly the structure of the beam (right). Both, the vertical and
    the horizontal axis are scaled by $R_0=Z_0=10^9$\,cm. A strong bow
    shock surrounds the jet, the cocoon, and the shocked progenitor
    gas.  A vigorous back flow from the head of the jet can be seen.
    Some vortices caused by Kelvin-Helmholtz instabilities are also
    indicated in the figure.
\label{fig:contour}}
\end{center}
\end{figure}

\begin{figure}
\begin{center}
\includegraphics[scale=0.5]{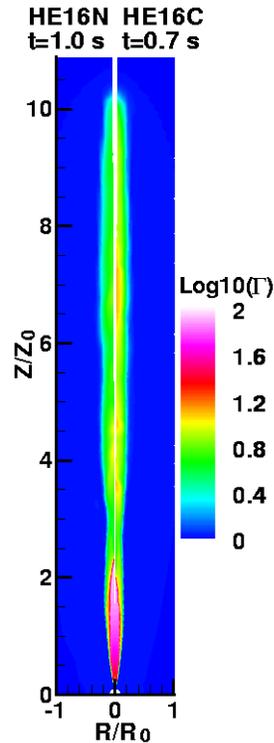}
\caption{Lorentz factor contours of two models (left: model HE16N at
  $t=1.0$\,s, and right HE16C at t$=0.7$\,s). Both, the vertical
  and the horizontal axis are scaled by $R_0=Z_0=10^9$\,cm.  The
  largest Lorentz factors are reached close to the polar axis
  where the first reconfinement shock appears.  Some weak reconfinement
  shocks can also be seen further downstream.
\label{fig:contour1_2}}
\end{center}
\end{figure}

Figure~\ref{fig:contour} shows a snapshot of the evolution of the
density of the model HE16N at $t=1.0$\,s, when the head of the jet is
still in the progenitor. The left panel of the
Fig.~\ref{fig:contour1_2} shows the Lorentz factor contour at the same
time shown in Fig.~\ref{fig:contour}.  The jet is well collimated both
inside of the progenitor and as it travels through the ISM.  The bow
shock develops close to the head of the jet and rises the pressure and
the temperature of the envelope region it sweeps up (in agreement with
the findings of \citealt{Mizuta04}).  It takes about 3.2\,s for the
jet to cross the progenitor, hence, the average propagation velocity
is $\sim 0.63c$.  The right panel of Fig.~\ref{fig:contour1_2} shows
the Lorentz factor of the model HE16C at $t=0.7$\,s, showing
  that the head of the jet in model HE16C propagates faster than model
  HE16N due to the lower density in model HE16C.

Although the outflow has a finite initial opening angle, the the beam
of the jet is almost parallel to the polar axis.  The kinetic energy
of the beam is dissipated when it crosses the reverse shock (i.e, the
Mach disc) at the head of the jet.  After the beam plasma is
decelerated at the Mach disc and its pressure is risen to a much higher
value than in the beam, it expands and flows back in a thick cocoon.
The high pressure of the cocoon is the responsible for the beam
collimation during the initial phase of propagation inside of the
progenitor star.  Also during this early stage of the evolution,
a strong backflow can be seen flanking the beam of the jet.  Some
vortices develop between the jet and the backflow caused by the growth
of Kelvin-Helmholtz modes. An schematic view of this process can be
seen in Fig.~5a of \citet{Mizuta06}.

The propagation of the jet inside the progenitor also drives a
cavity limited by a shroud whose density and pressure is larger
than in the cocoon.
The shroud is swept up by a reverse shock that results from
the interaction between the cavity and the progenitor
envelope. However, this reverse shock is not strong enough to rise
the temperature above the threshold in which nuclear reactions can
take place.

\begin{figure}
\begin{center}
\includegraphics[scale=.5]{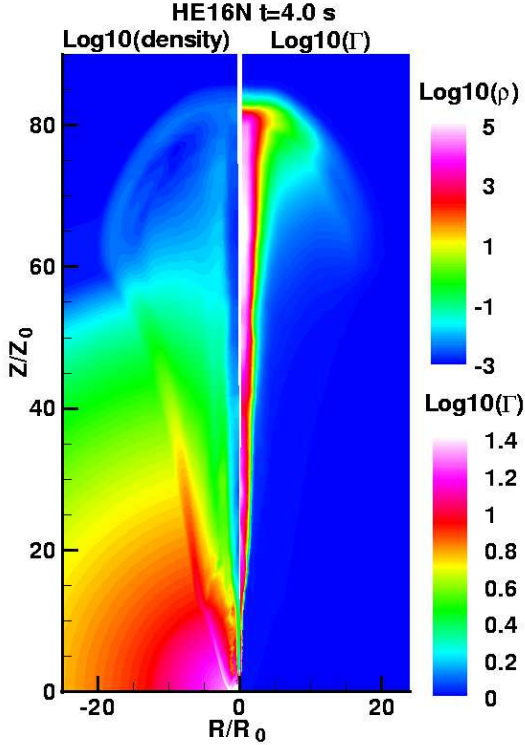}
\caption{Density and Lorentz factor contours of the model HE16N at
  $t=4$ s. Both, the vertical and the horizontal axis are scaled
    by $R_0=Z_0=10^9$\,cm. By this time, the head of the jet has
    already broken out of the progenitor surface, and the bow shock
    begins to expand into the ISM.
  \label{fig:contour2}}
\end{center}
\end{figure}

After the jet breaks out the progenitor surface, it proceeds to the
ISM, which is assumed to be rarefied for model HE16N.  In this phase,
cocoon is almost freely released into the ISM, because of the
negligible pressure of the external medium (Fig.~\ref{fig:contour2}).
In spite of the fact that the inertial confinement provided by the
stellar progenitor is lost in the ISM, by the time the jet reaches the
surface of the star, the beam has accelerated to $\Gamma \simmore 40$
(see Sect.~\ref{sec:acceleration}) and, thus, it has entered into a
ballistic regime, where lateral expansion is strongly
suppressed. Thereby, the jet remains well collimated as it
propagates through the ISM, and the half-opening angle of the beam
reaches only a few degrees. These collimation properties have
  been confirmed by means of numerical models with better resolution
  in the $\theta$-direction (see Appendix).

%%%%%%%%%%%%%%%%%%%%%%%%%%%%%%%%%%%%%%%%%%%%%%%%%%%%%%%%%%%%%%%%%%%%%%
%
\subsubsection{Acceleration to High Lorentz Factor}
\label{sec:acceleration}
%
%%%%%%%%%%%%%%%%%%%%%%%%%%%%%%%%%%%%%%%%%%%%%%%%%%%%%%%%%%%%%%%%%%%%%%

%
In this section we focus on the dynamics of the acceleration of the
flow to large Lorentz factors $\Gamma \simmore 100$, as requested by
the the standard fireball model \citep[e.g.,][]{Piran99}. Within such
a theoretical model, an initial release of thermal energy is later
converted into kinetic energy of the flow, as it expands into a dilute
and cold environment. On the basis of this model, a number of
numerical results have assessed that a jet, injected through a
pre-established nozzle, is able to drill its way through a collapsar
can reach an ultrarelativistic regime under likely inflow conditions
\citep[][ and this paper]{Zhang03,Zhang04,Mizuta06,Morsony07}. We
focus here on the details of the dynamical phase in which a
kinematically mildly relativistic jet ($\Gamma_\jj \sim 5$) speeds up
to the ultrarelativistic regime $\Gamma \simmore 100$ converting its
initial (thermal) energy $\epsilon_\jj /c^2\gg 1$ into kinetic
  energy.

\begin{figure}
\begin{center}
\includegraphics[angle=270,scale=0.38]{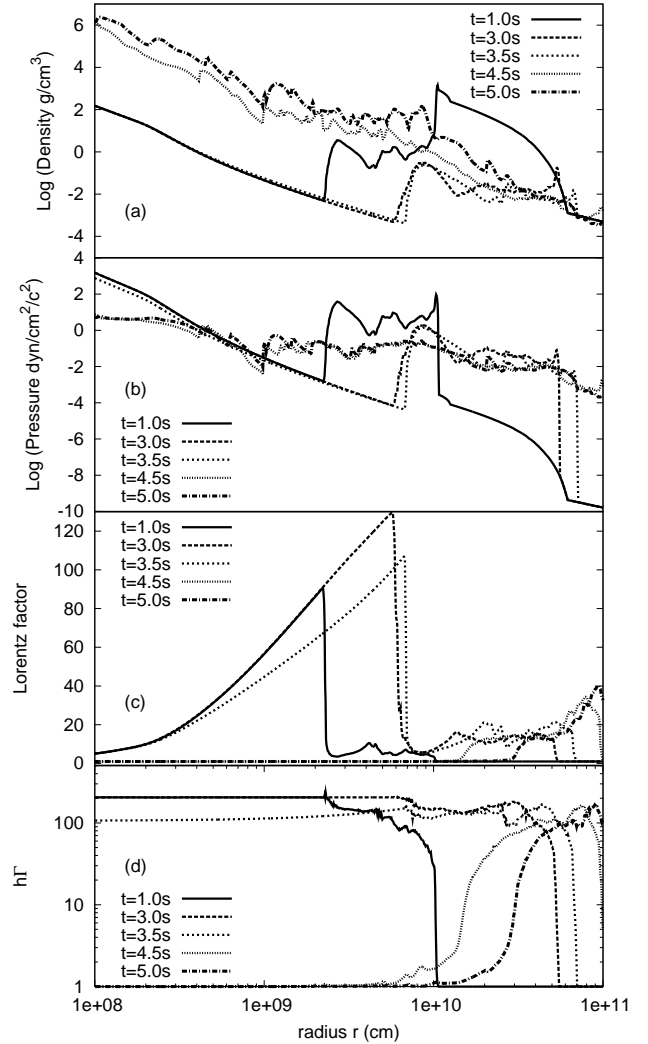}
\caption{Profiles of the rest-mass density (a), of the pressure
  (b), of the bulk Lorentz factor (c), and of $h\Gamma$
  (d) along the polar axis at times $t=$1.0, 3.0, 3.5, 4.5 and
  5.0\,s (see legends), corresponding to model HE16N.
  \label{fig:1dprofiles}}
\end{center}
\end{figure}

To drive the discussion, we take as a prototype case that of model
HE16N. The acceleration process of other models is very similar.
Figures~\ref{fig:1dprofiles}a-\ref{fig:1dprofiles}c show one
dimensional profiles of density, pressure, and Lorentz factor along
the polar axis, at different times. These profiles are
qualitatively similar to the ones shown in previous papers
\citep[e.g.,][]{Aloy02a,Morsony07}.  
During the injection the Lorentz factor increases linearly
(Fig.~\ref{fig:1dprofiles}c), whereas both the density and pressure
decrease as $r^{-3}$ and $r^{-4}$ (Fig.~\ref{fig:1dprofiles}a,
  b), respectively.  This is not unexpected, since the fluid expands
radially (almost freely) in that region, where there are no
shocks. Note that this result differs a bit from models where the
generation of the outflow is considered \citep{Aloy00,Aloy02a}.  If
jet injection conditions are set through a nozzle at a certain
distance to the center, the variability imprinted by the highly
dynamical generation of the jet is erased.  Clearly, this minimizes
the number of internal shocks in the outflowing jet. \cite{Aloy00}
show in their Fig.~2 that the outflow can accelerate to Lorentz
factors which are smaller than those attained in this work and in
others where jet injection conditions are assumed. This is because of
the modulation in the growth of the Lorentz factor imposed by the
development of Kelvin-Helmholtz modes in the course of the very early
outflow evolution \citep{Aloy02a}. Nevertheless, models which
consistently include the outflow generation (by thermal energy
deposition), can still accelerate to ultrarelativistic Lorentz factors
at large distance from the source. In such models the flow is kept
hotter than in the present ones because of the occurrence of internal
shocks already inside of the progenitor. However, since the outflow is
optically thick, the thermal energy is not radiated away, but
converted (later) into kinetic energy
\citep{Aloy00,Aloy02,Ghisellini07}.

One can notice that the maximum Lorentz factor is reached behind the
first recollimation shock in the jet (Figs.~\ref{fig:contour1_2}
  and \ref{fig:1dprofiles}c) coinciding with the location of the
minimum density in the beam of the jet (Figs.~\ref{fig:contour} and
\ref{fig:1dprofiles}a). Inside of the progenitor star the jet
  reaches a maximum Lorentz factor $\Gamma\simmore 100$ (see dashed
line in Fig.~\ref{fig:1dprofiles}c corresponding to $t=3\,$s).
We note that the quick acceleration to large Lorentz factor happening
in the present models is a direct result of the fact that the flow is
injected into a finite opening angle.  Differently, \cite{Mizuta06},
who injected the jet parallel to the polar axis of the progenitor, did
not find such a fast acceleration.  In the later case, the occurrence
of recollimation shocks much closer to the injection nozzle prevents
the development a freely expanding, unshocked jet extending so far
away as in the models of this work.  Indeed, such internal,
recollimation shocks are the responsible of the confinement of the jet
\citep{Komissarov97,Komissarov98,Mizuta04,Mizuta06,Mimica08}.
Nonetheless, the head propagation speed is very similar in
\cite{Mizuta06} as it is found here and, hence, also the crossing
time of the progenitor by the jet.  In \cite{Mizuta06} an acceleration
of the jet to $\Gamma \simmore 100$ at larger distances from the
source happens (by thermal-to-kinetic energy conversion), but it still
takes place inside of the progenitor star. This resembles more the
acceleration process of outflows generated self-consistently from the
inner engine as pointed out above.

The absolute maximum Lorentz factor ($\Gamma_{\rm max}\sim 130$) is
attained by the time we begin to reduce the luminosity of the inflow
(at $t=3$\,s, see \S~\ref{sec:numerics}). By this time, the head of
the jet has already broken up the surface of the progenitor. This
numerical value is smaller than the maximum one which could be
potentially reached, provided the conditions of the injected outflow,
according to which $\Gamma_{\rm max}\sim \Gamma_{\jj,0} h_\jj \simeq
205$, which is an indication of the maximum
potentially reachable Lorentz factor, along a flow line, if all
thermal energy were converted to kinetic energy.  
After the energy
deposition is switched off ($t=3\,$s), the region previously
occupied by the unshocked beam of the jet (up to the first
recollimation shock of the beam; Fig.~\ref{fig:1dprofiles}c) begins
to shrink and to incorporate mass from its lateral boundaries
  (c.f., rest-mass density profiles between times 3.5\,s and 5.0\,s in
  Fig.~\ref{fig:1dprofiles}a up to $r \sim 8\times10^9\,$cm). This
  fact reduces progressively both the beam Lorentz factor
  (Fig.~\ref{fig:1dprofiles}c) and the ability of the beam fluid to
  reach large asymptotic values of $\Gamma$
  (Fig.~\ref{fig:1dprofiles}d). Thereby, if the jet is injected with
a finite half-opening angle, the ultrarelativistic part of the outflow
(i.e., that where $\Gamma \simmore 100$) may persist only if the
activity of the engine does not cease.

%%%%%%%%%%%%%%%%%%%%%%%%%%%%%%%%%%%%%%%%%%%%%%%%%%%%%%%%%%%%%%%%%%%%%%
%
\subsection{Angular Energy Distribution}
\label{sec:dEdOmega}
%
%%%%%%%%%%%%%%%%%%%%%%%%%%%%%%%%%%%%%%%%%%%%%%%%%%%%%%%%%%%%%%%%%%%%%%

The evolution of our models has been followed until the head of the
jet reaches the outermost radial computational boundary at $r_{\rm
  max}=10^{11}\,$cm. At such distances, the outflow is still optically
thick and, therefore, radiation cannot escape freely. The jet and
the surrounding cocoon, past a transient phase, that happens after
the jet breaks out of the stellar surface, enter into a quasi
self-similar phase, where the properties of the outflow roughly scale
with the distance of propagation and become almost time
independent. This fact allows us to extrapolate the properties of the
jet from distances $r \simeq 10^{11}\,$cm to the typical distances
where the afterglow will take place (namely, $r\simeq
10^{15}\,$cm). Thus, we can roughly estimate the angular energy
distribution per unit solid angle ($dE/d\Omega$) that can be
potentially emitted at the afterglow phase.

In order to derive $dE/d\Omega$ we have to make several
assumptions. First, we assume that a fix fraction (the same
everywhere) of the total energy (internal plus kinetic) will be
converted into electromagnetic radiation. Basically, this assumption
is equivalent to state that the angular profile of the observed
non-thermal radiation is simply a scaled version of the total energy
angular profile. Certainly, this is a rough approximation, since the
non-thermal radiation from $\gamma$-rays to radio frequencies will be
produced by synchrotron (and, perhaps, inverse Compton) processes of
particles accelerated at shocks (or, maybe, along the jet boundary
layer; e.g., \citealt{Aloy06}). Obviously, there are shocks of very
different properties in the ultrarelativistic beam and in the cocoon
and, thus, we may expect somewhat different conversion efficiencies of
the outflow energy into radiation in the beam and in the
cocoon. Finally, we assume that the angular energy distribution is
{\it frozen-in} by the time when the head of the jet reaches the outer
computational boundary.  As commented above, our models evolve almost
self-similarly a bit after they break out of the stellar surface, and
therefore, we expect only a minor time evolution of the angular
profiles of $dE/d\Omega$.

We point out that the procedure we use to estimate $dE/d\Omega$
 differs from that of \cite{Morsony07}, who derived their
$dE/d\Omega$ profiles from the time integration of the energy flux
trough a certain radius. Under the hypothesis of self-similar
evolution, this is equivalent to integrate, along the radial
direction, the energy density of our models (by the time they
  reach $r_{\rm max}$) as follows,
\begin{eqnarray} {dE \over d\Omega} (\theta)\equiv 
\sum_{k=-N_\theta}^{N_\theta} \sum_{i=i_*}^{N_r}
  \left(1-\beta^r_{ik}\over 1-\beta^r_{ik}
    \cos{(\theta-\theta_{k})}\right)^3\nonumber \\
\times (\rho_{ik} h_{ik}
  {\Gamma_{ik}} ^2-p_{ik}-\rho_{ik}\Gamma_{ik}){r_i}^2 \Delta r_i,
\label{eq:dEdOmega}
\end{eqnarray}
\noindent
where the subscripts $i$ and $k$ are associated to the spherical grid
coordinates $r_i$, and $\theta_k$, respectively, and $\beta^r$ and
$\theta$ are the radial velocity in units of $c$ and the observer's
viewing angle (measured from the jet axis), respectively.  The
expression~(\ref{eq:dEdOmega}) includes the radiation contributions
coming from regions outside of the line of sight (see
\citealt{Janka06}).  The summation in the radial direction runs from
the surface of the progenitor, located at $r=r_*$ or, equivalently,
$i=i_*$, to the outermost boundary.  The summation in the
  azimuthal angle runs from $\theta=-90^{\circ}$ to
  $\theta=90^{\circ}$ (note that due to the assumed axial symmetry, we
  can copy the computed data of the quadrant
  $0^{\circ}<\theta<90^{\circ}$ to the quadrant
  $-90^{\circ}<\theta<0^{\circ}$).  In order to avoid accounting for
subrelativistic regions, which will not contribute to the afterglow
energetics, we exclude the contributions of numerical cells where
$v_r<0.7c$ and $h\Gamma<4$ in the
expression~(\ref{eq:dEdOmega}).\footnote{Note that numerical cells
  where $h\Gamma<4$ have the potential to accelerate, at most, to
  $\Gamma\sim 4$. In actuality, the asymptotic Lorentz factor of such
  parcels of fluid, will be much smaller, since they will decelerate
  as they incorporate mass from the external medium.}

\begin{figure}
\begin{center}
\rotatebox{270}{\includegraphics[scale=0.35]{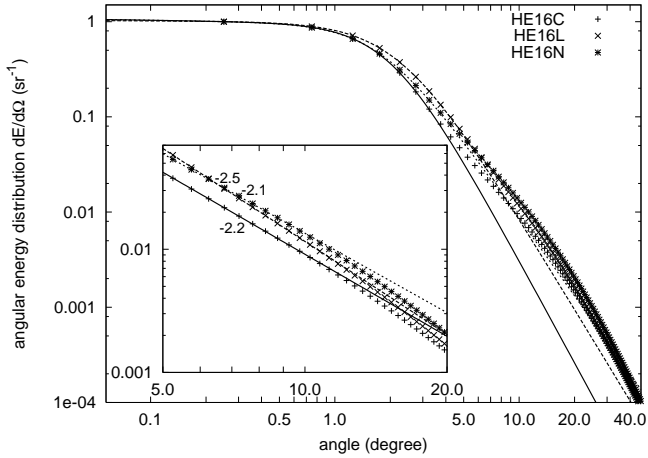}}
\caption{Angular energy distribution of models, HE16C, HE16L, and
    HE16N, when the head of the jet reaches the outer computational
  boundary.  With lines we also show the fitting functions to SBPs
    (Eq.~\ref{eq:fitting}) for each model.  The solid, dashed, and
    dotted lines are the fitting functions of Eq.~\ref{eq:fitting}
    corresponding to models HE16C, HE16L, and HE16N, respectively.
    The inset displays a zoom of the angular range
    $5^\circ<\theta<20^\circ$, where simple power law fitting
    functions are overplotted, along with the corresponding values of
    the power law indices. This angular region is dominated by the
    contribution of the expanding cocoon.
    \label{fig:distri1}}
\end{center}
\end{figure}

The absolute value of the observed $dE/d\Omega(\theta)$ along every
radial direction forming an angle $\theta$ with the polar axis
depends, among other things, on two parameters whose exact value is
not well constrained, neither by observations nor by the present day
theory.  These are (i) the efficiency of energy conversion to
radiation, and (ii) the total energy injected. Therefore, we will
show only the angular profiles of $dE/d\Omega(\theta)$ normalized to
the maximum value $\left. dE/d\Omega(\theta)\right|_{\rm max}$ found
for each model. Figure~\ref{fig:distri1} shows the normalized angular
energy distributions corresponding to models HE16C, HE16L and HE16N,
which are prototypes of the types L, M and H, respectively. In the
same figure we overplot fits to the normalized $dE/d\Omega(\theta)$
profiles. The fitting function is a smoothly broken power law (SBP) of
the form,
\begin{eqnarray}
\label{eq:fitting}
F(\theta)=2^{-1/n}A\left[\left({\theta\over \theta_0}\right)^{\alpha_l n}
+\left({\theta\over \theta_0}\right )^{\alpha_h n}\right]^{1/n},
\end{eqnarray}
where $A$ is the value of the function $F$ at $\theta=\theta_0$,
$\theta_0$ is the angular location of the break point between
the prebreak and postbreak power-laws, whose slopes are $\alpha_l$ and
$\alpha_h$, respectively, and $n$ is a numerical factor that controls
the sharpness of the break. Note that the maximum value of $F$ occurs
at
\begin{equation}
  \theta_{\rm max}= \theta_0 \left(-\frac{\alpha_h}{\alpha_l}
  \right)^{1/[(\alpha_l - \alpha_h) n]} ,
\label{eq:theta_max}
\end{equation}
when $\alpha_l \alpha_h<0$.
Otherwise, if $\alpha_l<0$ and
$\alpha_h<0$, the function diverges as $\theta \rightarrow 0$.

By inspection of Fig.~\ref{fig:distri1}, the angular energy
distributions are remarkably well fitted by the function of
Eq.~(\ref{eq:fitting}) in the interval $0< \theta \lesssim
3.4^{\circ}$, i.e., in the angular region occupied by the beam of
  the jet.
At smaller latitudes ($5^{\circ} \lesssim \theta \lesssim
8^{\circ}$) the model data separates from the fitting function and
presents systematically larger values than the latter. Indeed, the
data in such an interval can be well fitted by a simple power law,
with a slope in the range $[-2.1, -2.5]$ 
(see the inset in Fig.~\ref{fig:distri1}).  The deviation from the SPB
function in this angular range is due to the contribution of the
expanding, mildly relativistic cocoon. This cocoon
  contribution shows up more clearly (the energy distribution tends to
  flatten in the range $\theta> 5^\circ$) if we lower the thresholds
  on the values of $h\Gamma$ and $v_r$ used to compute the angular
  energy distribution of each model. However, as we argued before,
lowering these thresholds
too much will
catch up contributions from
  numerical cells whose asymptotic Lorentz factor is too small to
  account for typical afterglow.

\begin{figure}
\begin{center}
\rotatebox{270}{\includegraphics[scale=0.35]{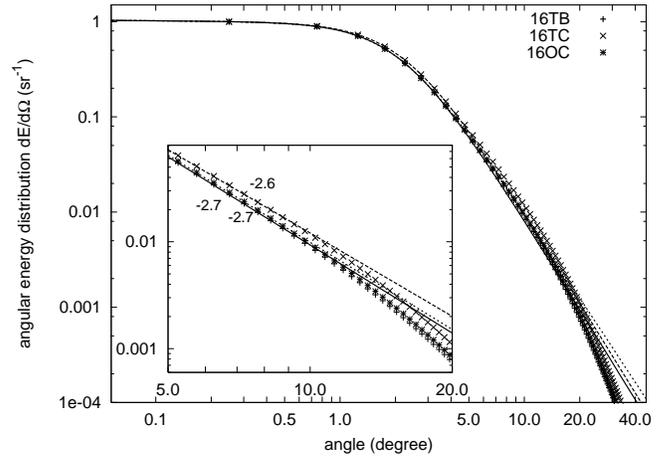}}
\caption{Same as Fig.~\ref{fig:distri1}, but for the models, 16TB,
  16TC, and 16OC.
  \label{fig:distri2}}
\end{center}
\end{figure}

Low metallicity models 16TB, 16TC, and 16OC, are also well fit by the
function of Eq.~(\ref{eq:fitting}) in roughly the same interval as the
solar metallicity models (Fig.~\ref{fig:distri2}).  The values of the
fit parameters are comparable to those of type-H models, with
which they share a very similar progenitor mass ($\sim 15 \Msun$).
However, in these models, the cocoon contribution, which can be fit by
a simple power-law with a slope $\sim -2.6$ between $5^{\circ}
\lesssim \theta \lesssim 8^{\circ}$ (Fig.~\ref{fig:distri2} inset)
shows a faster decay of $dE/d\Omega$ than type-H models for
$\theta>8^\circ$. The reason for this difference is the much deeper
density drop of low-metallicity models close to the star surface
(Fig.~\ref{fig:1dradialmass2}) compared with type-H models
(Fig.~\ref{fig:1dradialmass}). The density of low-metallicity models
in the region $3\times10^{10}\,$cm\,$\lesssim r\lesssim
4\times10^{10}\,$cm is $\sim 100$ times smaller than in type-H
models. Hence, the beam of jets in such low-metallicity progenitors
becomes much more ballistic than the corresponding beams of jets in
the type-H group. Since more ballistic beams reduce the sideways
expansion of their cocoons, this explains that the angular energy
distribution in low-metallicity stars is more narrowly concentrated
than in solar-metallicity, type-H progenitors.

\begin{figure}
\begin{center}
\includegraphics[scale=0.34]{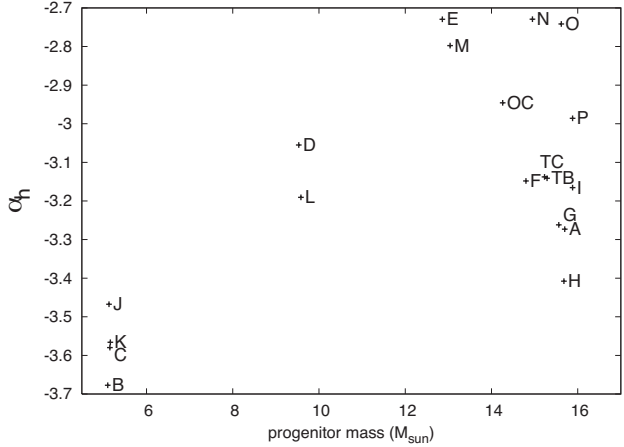}
\caption{Dependence of the index $\alpha_h$ on the total mass
of the progenitor. We identify the models by the last letter in
the model name, e.g., the label 'A' stands for model HE16A.
The labels 'TB', 'TC', and 'OC' stand for the models 16TB, 16TC, and
  16OC, respectively.
\label{fig:mass_al}}
\end{center}
\end{figure}

In order to show more clearly the existence of correlations between
the properties of the progenitor star and the $dE/d\Omega$
distribution, we show in Fig.~\ref{fig:mass_al} the dependence of the
postbreak slope $\alpha_h$ and on the stellar progenitor mass
$M$. There exists a correlation between $\alpha_h$ and $M$, such that
the slope of lighter progenitors is steeper than that of heavier
ones. There is roughly a linear dependence of $\alpha_h$ on $M$, which
displays a relatively large dispersion. The reason for the dispersion
being that for very similar values of the total progenitor mass, the
rest-mass density radial profiles can be appreciably different (see,
e.g., Figs.~\ref{fig:1dradialmass} and \ref{fig:1dradialmass2}). This
is particularly true in heavy progenitor models (including type-H and
low metallicity models). For the prebreak slope $\alpha_l$ we
  find no obvious correlation with the progenitor mas, but in all the
  models considered here is very small ($\alpha_l\simeq 0$;
  Tab.~\ref{tab:progenitor}).

\begin{figure}
\begin{center}
\includegraphics[scale=0.35]{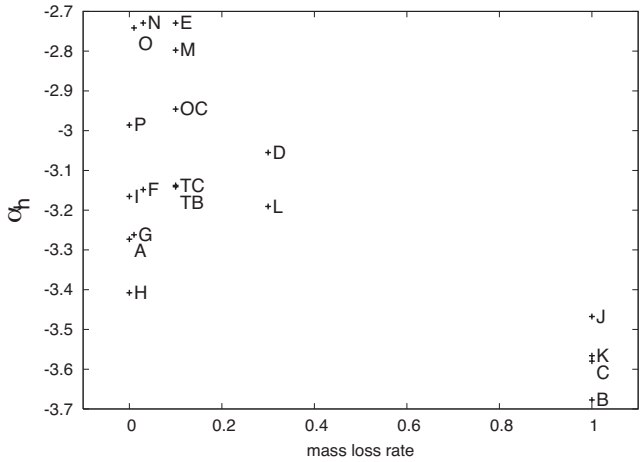}
\caption{Dependence of the index $\alpha_h$ on the mass loss rate
  in units of the {\em typical} mass loss rate considered
by \cite{Woosley06}.
The naming convention is the same as in  Fig.~\ref{fig:mass_al}.
\label{fig:massloss_al}}
\end{center}
\end{figure}

We have also investigated the dependence of the slope $\alpha_h$ and
$\alpha_l$ on the mass loss rate $\dot{M}$ assumed in models of
\citet{Woosley06} (see Tab.~\ref{tab:progenitor}).
Figure~\ref{fig:massloss_al} shows that there exist a good
correlation between $\alpha_h$ and $\dot{M}$, while $\alpha_l$
  seems to be independent of $\dot{M}$. The $\alpha_h - \dot{M}$
  correlation tells us that, models with a larger mass loss rate
posses a steeper slopes. This is not surprising considering the
previously found correlation between $\alpha_h$ and $M$, since the
stellar progenitor mass is mostly determined by the amount of mass
lost in the form of winds during the latest stages of its evolution.

We have not found any other good correlation between the fit
parameters (other than $\alpha_h$) and the gross properties
of the progenitors (radius, average density, total angular momentum,
rotation period, mass of the iron core, etc.).

\begin{figure}
\begin{center}
%\rotatebox{270}{\includegraphics[scale=0.34]{fig11_2.eps}}
\rotatebox{270}{\includegraphics[scale=0.34]{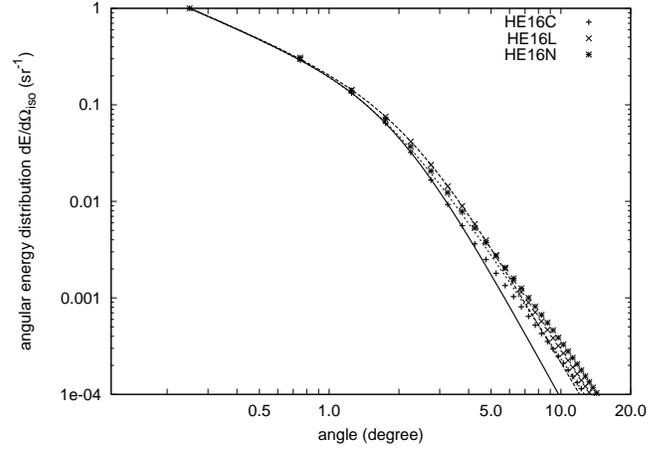}}
\caption{Equivalent isotropic angular energy distribution of
  the jets when the head of the jet reaches the outer computational
 boundary. The models shown are HE16C, HE16L, and HE16N, along with
 their corresponding fitting functions (with lines).
\label{fig:distri3}}
\end{center}
\end{figure}

\begin{figure}
\begin{center}
\includegraphics[scale=0.34]{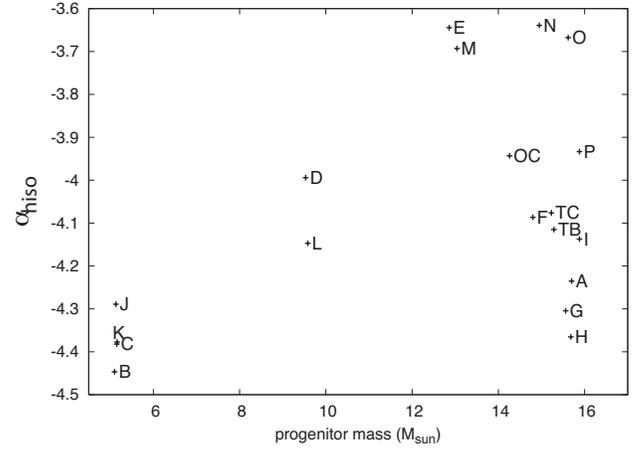}
\caption{Same as Fig.\ref{fig:mass_al} but for the equivalent
  isotropic energy distribution.
\label{fig:mass_al2}}
\end{center}
\end{figure}

We shall note that the angular distribution $dE/d\Omega$ is not
directly observable. Instead, the isotropic equivalent angular energy
per solid angle $dE/d\Omega|_{\rm iso}$ can be detected.
Figure~\ref{fig:distri3} shows the equivalent isotropic angular energy
distribution for prototype models belonging to the types L, M and H,
normalized to the value of the distribution at $\theta=0.25^\circ$.
As can be seen from Fig.~\ref{fig:distri3} (see also
Tab.~\ref{tab:progenitor}), the values of \aliso\, are
also negative for the $dE/d\Omega|_{\rm iso}$, while the
corresponding values $\alpha_l$ are close to zerofor $dE/d\Omega$
(approximately, \aliso$\simeq\alpha_l -1$). This happens because of
the small value of the solid angles close to the symmetry axis, which
makes systematically larger the higher latitude values of
$dE/d\Omega|_{\rm iso}$ than those of $dE/d\Omega$. The
values of \ahiso\, are systematically smaller than the respective
$\alpha_h$ values (roughly, \ahiso$\simeq\alpha_h -1$ in our
  standard resolution runs). Given the tight relations of the slopes
in the $dE/d\Omega$ and $dE/d\Omega|_{\rm iso}$ distributions, it is
not surprising to find that there exists a good correlation between
\ahiso\, and $M$ (Fig.~\ref{fig:mass_al2}) 
which follows the same qualitative trend as the correlation
between $\alpha_h$ and $M$.

\begin{figure}
\begin{center}
\includegraphics[scale=0.5]{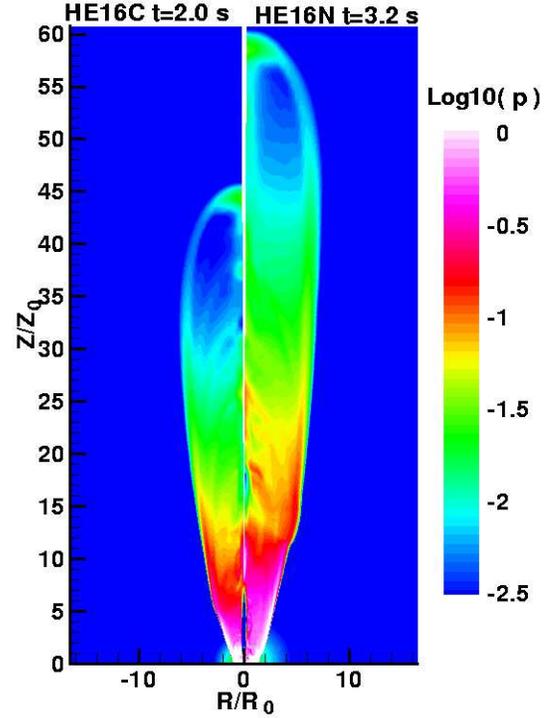}
\caption{ Pressure contours of models HE16C (left: at $t=2.0$\,s) and
  HE16N (right: at $t=3.2$\,s), when the head of the jet reaches
  progenitor surface in each model, i.e., at shock break out.
  Both, the vertical and the horizontal axis are scaled by
    $R_0=Z_0=10^9$\,cm. The radii of the progenitors are $4.73\times
  10^{10}$\,cm and $6.13\times 10^{10}$\,cm for the models HE16C and
  HE16N, respectively.  At the same distance $Z/Z_0$ from the
  injection nozzle, the pressure in both the jet and in the cocoon is
  higher for model HE16N than for model HE16C. This fact explains the
  larger lateral expansion after shock break out in model HE16N than
  in model HE16C.
  \label{fig:comparison}}
\end{center}
\end{figure}

\begin{figure}
\begin{center}
\rotatebox{270}{\includegraphics[scale=0.32]{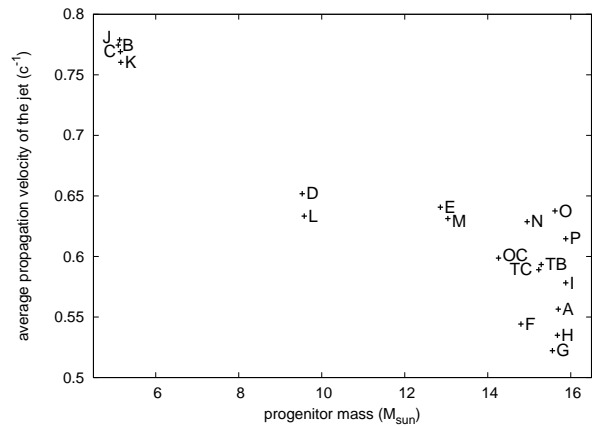}}
\caption{Dependence on the progenitor mass of the average
  propagation velocity of the jet in the star. The average
  propagation velocity is defined as the ratio between the
  progenitor radius $r_*$ and the time it takes the head of the jet
  to propagate up to $r_*$. The labeling convention is the same
  as in Fig.~\ref{fig:mass_al}.
\label{fig:average_velo}}
\end{center}
\end{figure}

\begin{figure}
\begin{center}
\rotatebox{270}{\includegraphics[scale=0.32]{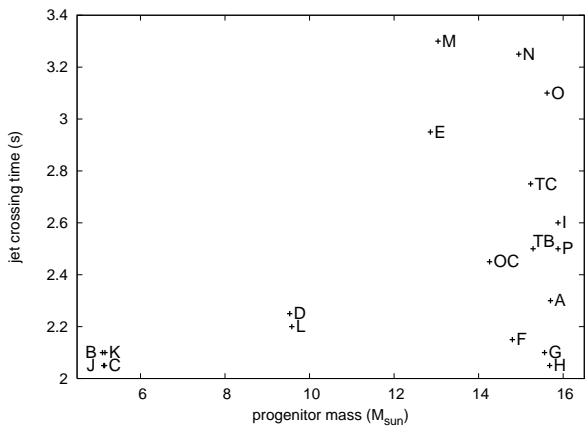}}
\caption{Dependence on the progenitor mass of the time
employed by the jet to reach the progenitor
surface. The labeling convention is the
  same as in Fig.~\ref{fig:mass_al}.
\label{fig:time_cross}}
\end{center}
\end{figure}

To sum up our findings so far, we realize that the values of the
parameters of the fit function (Eq.\ref{eq:fitting}) are chiefly
correlated with the mass of the progenitor. In type-L progenitors, the
velocity of propagation of the jet is larger than in more massive
models, which results in jets developing more massive and hotter
cocoons in type-M and type-H models than in light
progenitors\footnote{Since the injected mass flux is the same in all
  models, jets which take longer to reach the stellar surface (i.e.,
  jets with small propagation speed) are more massive and, hence their
  cocoons are also heavier. On the other hand, the slower jets develop
  stronger reverse shocks than the faster ones, because the injected
  fluxes of mass, momentum and energy are the same in all our
  models. Thus, the jet matter is compressed more in the former than
  in the latter case, which explains why the matter in the cocoon is
  hotter in jets with smaller propagation speeds.}. This fact is
evident by looking at Fig.~\ref{fig:comparison}, where the pressure
in the cocoon of the model HE16N is higher than that in the
model HE16C in the course of the propagation of the jet up to the
progenitor surface. The differences in the propagation velocity of the
injected jets are set, to a large extend, by the differences in the
average density between progenitors of different mass. The less
massive progenitors of our sample (type-L models) tend to have the
smaller average densities (see, e.g., Fig.~\ref{fig:1dradialmass},
where model HE16C displays a smaller density than model HE16N at every
radial point). Consistently, jets propagating in type-L progenitors
are faster as can be observed in Fig.~\ref{fig:average_velo}. However,
the time needed to reach the progenitor surface presents a dependence
with the progenitor mass with a much larger scatter
(Fig.~\ref{fig:time_cross}). Particularly, models HE16A,
  HE16F, HE16G and HE16H, all of which belong to the type-H
group, display a progenitor crossing time comparable to that of the
jets in the type-L group (although its average propagation velocity is
comparable to that of models of type-H). That's the reason why 
  these models appear as {\em outliers} in the correlations between
the slope ${\alpha_h}$ and
the progenitor mass (Figs.~\ref{fig:mass_al} 
and with the mass-loss rate
(Fig.~\ref{fig:massloss_al}).
Since more massive progenitor stars yield hotter cocoons, once the
cocoon is erupted through the stellar surface, it undergoes a larger
lateral expansion (compare models HE16C and HE16N in the lower panels
of Fig.~\ref{fig:1dradialmass}), i.e., the energy carried by the jet
spreads towards lower latitudes. This explains why the energy per
solid angle is more concentrated towards the axis when the jet crosses
a low mass progenitor.

\begin{figure}
\begin{center}
\rotatebox{270}{\includegraphics[scale=0.34]{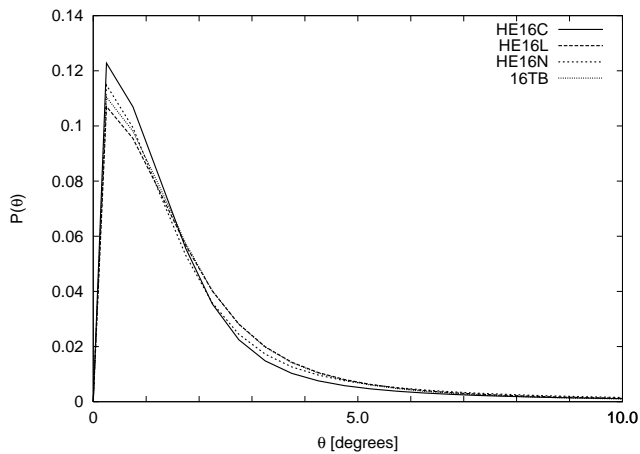}}
\caption{Detectability of the models HE16C, HE16L, HE16N and 16TB.
\label{fig:detecta}}
\end{center}
\end{figure}

If we assume that the detectability of an event, for an observer
looking such event at a certain viewing angle ($\theta$), is
proportional to $dE/d\Omega|_{\rm iso} (\theta)$ (see, e.g.,
\citealt{Janka06}), it turns out that $\theta$ should be rather small
($\simless 2^\circ$; Fig.~\ref{fig:detecta}) to observe and event
produced in collapsar progenitors (note that the flanks of
$dE/d\Omega|_{\rm iso}$ distribution are quite steep and, thus, it is
very unlikely to detect events which are not directly pointing towards
the observer). Jets  produced in type-L progenitors exhibit
narrower observability profiles than those injected in more massive
starts (c.f., compare the profiles of models HE16C and HE16N in
Fig.~\ref{fig:detecta}). Therefore, it is more unlikely to detect
off-axis events produced in light progenitors than in more massive
ones. Alternatively, we may state the the lower degree of
  collimation of relativistic jets in high-mass stars results in a
  higher probability of observing an event from a high-mass progenitor
  than a low-mass progenitor.  We also find that the metallicity has
little influence on the observability, because for similar progenitor
masses, solar-metallicity models and low metallicity ones display
almost identical observability angular profiles
(Fig.~\ref{fig:detecta}).

  We also notice the very different observability profiles of jets
  produced in collapsars and jets produced in remnants of mergers of
  compact objects. In the latter case, a non-negligible observability
  is obtained in the viewing angle interval $2^\circ \simless \theta
  \simless 13^\circ$, because of the larger half-opening angles of the
  outflows originated in progenitors of short GRBs (see Fig.~2 of
  \citealt{Janka06}). Another relevant difference is that
  collapsar-jets show a detectability with a fast rise close to
  $\theta=0$, followed by a shallower decay beyond the most probable
  detection angle $\sim 0.5^\circ$. In contrast, most jets produced in
  merger remnants tend to decay very abruptly beyond the most likely
  observing angle, and have an observability rise much more moderate
  than that of collapsar-jets. The reason for this difference is the
  much shallower decay of the isotropic equivalent energy at the
  flanks of the beam in collapsar-jets than in jets from merger
  remnants.

\section{CONCLUSIONS AND DISCUSSION}
\label{sec:conclusions}

In this paper we explore the relation of the dynamical properties of
ultrarelativistic jets generated in collapsars with the properties of
the progenitor stars in which such jets propagate. We have
particularly focused on the correlations that exist between the
angular profile of the energy per solid angle (as seen in the
afterglow phase) and the properties of the progenitors. Along the way,
we have pointed out which is the relevance of the fact that our
numerical models are set up with a finite injection half-opening
angle.

Using a non-zero injection angle affects the way in which the
conversion of internal-to-kinetic energy takes place. If the flow is
injected parallel to the polar axis, the development of reconfinement
shocks happens closer to the injection nozzle than if the flow is
injected radially within a cone of finite half-opening angle. When the
recollimation shock occurs far away from the nozzle, the unshocked
beam flow accelerates along a larger distance in a rarefaction that
precedes such shock. Thus, the beam reaches there larger Lorentz
factors than if the jet is injected parallel to the polar axis. The
dissipation in cross shocks acts by simply recycling part of the
kinetic energy of the outflow into thermal energy. The thermal
energy is not lost, since the jet propagation is roughly adiabatic
inside of the progenitor (radiation losses are negligible
there). Instead, this thermal energy can be further converted into
kinetic energy at larger distances. This process may happen several
times before the outflow becomes transparent and radiation can freely
escape. This explains why, in spite of the differences in the beam
dynamics, all models (independent of the injection half-opening
angle; c.f., \citealt{Mizuta06}) develop a roughly similar
propagation speed and, by the time they reach the head of the jet, the
gross properties of the outflow are similar.

We have estimated the angular distribution of energy per solid angle
in the afterglow phase by extrapolation of the state of our models
when they reach a distance of $\sim 10^{11}\,$cm. This extrapolation
relies on the fact that the jets develop a rough self-similar behavior
soon after they emerge from the progenitor surface. Our results show
that the equivalent isotropic energy per solid angle $dE/d\Omega|_{\rm
  iso}$ is only partly consistent with that of
\cite{Lazzati05}. However, the results in this paper do appear
  to be consistent with previous numerical simulations such as those
  of \cite{Zhang04} and \cite{Morsony07}. \cite{Lazzati05} obtain
that the angular energy distribution displays a relatively flat core
which is flanked by a region where $dE/d\Omega|_{\rm iso} \propto
\theta^{-2}$. Our results show that the core of the distribution
(close to $\theta=0^\circ$) is not flat (but decays as
  $\theta^{-1}$) and that the energy per solid angle decays much
faster than $\theta^{-2}$ (it does it as $\theta^{\kappa}$,
  with a value of $\kappa\simless -3.6$ depending on the mass of the
  progenitor; see below). We can fit the $dE/d\Omega|_{\rm iso}$-data
with SBP functions up to $\theta\simless 3.4^\circ$. At
smaller latitudes, a simple power-law with a slope close to $-2.6$
fits better the data. In this region ($4.5^\circ < \theta < 8^\circ$),
the cocoon contribution is the dominant (at smaller values of
$\theta$, the beam of the jet dominates the energetics), and we find
there the best consistency with the model of \cite{Lazzati05}.

We have correlated the properties of the angular $dE/d\Omega|_{\rm
  iso}$ distribution of the jet with the fundamental parameters
of the progenitor star in which the jet has propagated. We
find that the shape of the distribution is mostly influenced by the
mass of the progenitor. When the mass of the progenitor is small
($M\sim 5 \Msun$), because of the occurrence of large mass losses due
to winds in the latest stages of the star's evolution, then
$dE/d\Omega|_{\rm iso}$ decays faster with $\theta$ than if the mass
of the progenitor is large ($M\sim 15 \Msun$). We find that the reason
for this behavior is that the average density of the progenitors tends
to grow (approximately) with the mass. This means that the average jet
propagation speed inside of the star is smaller, the larger is the
mass. A smaller jet propagation speed results into thicker and hotter
cocoons, since we fix the same mass, momentum and energy fluxes at the
nozzle for all models. Also the beam of the jet in the more massive
progenitors is wider. This is the reason why low mass progenitors
develop narrower $dE/d\Omega|_{\rm iso}$ profiles than high mass
ones. The difference in the collimation of the energy distributions
resulting from low and high mass progenitors has a direct influence on
the number of observed events. We expect to see more events produced
in heavy progenitors than in low mass ones.

One could question whether the correlation that we have found between
the mass of the progenitor and the width of the $dE/d\Omega|_{\rm
  iso}$ is an artifact of our numerical set up. We are fixing the
luminosity of the jet to be the same independent of the progenitor
mass. However, progenitors of different mass may develop central
engines which release different power. A good proxy of the power of
the central engine is the mass of the iron core of the progenitor. One
may expect that the collapse of more massive iron cores results into
larger central compact objects. If the power released by the central
engine is dominated by the size of the central compact object, then
models with more massive iron cores could release a larger power than
models with low mass cores. Then, how do we justify our numerical
assumption that the power injected in the jet is roughly independent
of the progenitor's mass?. In support of our point we argue that,
first, according to \cite{Woosley06} data, there is not a one to one
correlation between the mass of the iron core and the mass of the
progenitor and, second, the iron mass varies by less than a 30\% in
all the models considered here, while the total mass can be different
by a factor of 3. Thus, within the simplifications we do in our
models, the assumption of a common luminosity independent of the mass
of the progenitor is justified.

Irrespective of these two arguments, if heavier progenitors would
result into more luminous central engines, we shall point out that
this trend will also result into wider jets and cocoons. This is a
result early pointed out by \cite{Aloy00}, were it was shown that,
taking the same progenitor, but increasing the luminosity of the
central engine by a factor of 10, results into thicker cocoons than in
cases in which the luminosity is more moderate. The reason being that
the release of a large power triggers large amplitude Kelvin-Helmholtz
instabilities at the basis of the outflow, which transfer a large
fraction of the momentum of the beam to a thicker shear layer between
the beam and the cocoon. Effectively, this process widens the cross
sectional area of the beam, reducing its propagation speed and,
consistently inflating larger cocoons.

We have also found, that comparing progenitors of similar mass,
the metallicity of the star has a small impact on the extrapolated
$dE/d\Omega|_{\rm iso}$ profiles. The collimation of the jet is
similar regardless of the stellar metallicity. However, the cocoon
is more narrowly collimated in low-metallicity stars, because of the
large density drop close to their surfaces. This reduced density
makes the jets in low-metallicity stars much more ballistic, once
they break out the stellar surface, than in solar-metallicity
progenitors. Unfortunately, this difference might not be
observable, since it happens in regions where the energy per solid
angle is much smaller than at the jet core (unless orphan afterglows
could be detected; see, e.g., \citealp{Totani02,Rossi08}).

Finally, we have found significant differences between the
$dE/d\Omega|_{\rm iso}$ profiles of collapsar-jets and those of jets
produced in merger remnants (i.e., between the angular energy
distribution of jets associated to long and to short
GRBs). Collapsar-jets are more narrowly collimated than jets of
merger remnants, and the decay of $dE/d\Omega|_{\rm iso}$ beyond the
central flat core is much steeper in the latter than in the
former. This intrinsic difference manifest itself as a larger chance
of detectability of jets from merger remnants at viewing angles up
to $\theta \sim 12^\circ$, while, on the other hand, collapsar-jets
could hardly be seen off-axis.

\acknowledgments

We would like to thank A. Heger for his kindness to allow us to use
his progenitor models for this study.  We would like to thank
  Dr. Hanawa for his helpful comments.  We also thank the referee for
  his/her useful comments that helped to improve this paper.  This
work is partly supported by the Grants-in-Aid of the  Japanese
Ministry of Education, Science, Culture, and Sport (19540236,
20041002, 20340040 A.M.), the grants of the Spanish Ministry of
Science and Innovation CSD2007-00050 and
  AYA2007-67626-C03-01. M.~A.~A. is a Ram\'on y Cajal fellow of the
Spanish Ministry of Education and Science.  The calculations were
carried out on a NEC SX8, at the Cybermedia Center, Osaka University,
on a NEC SX8, at Yukawa Institute for theoretical physics, Kyoto
University, on the Space Science Simulator (NEC SX6), at the Japan
Aerospace Exploration Agency, and on a NEX SX9, at the Center for
Computational Astrophysics, National Astronomical Observatory of
Japan.

\begin{appendix}
\section{On the choice of numerical resolution}
The main purpose of this Appendix is to justify the choice of
numerical resolution that we have used in the main body of this
paper. We have picked up three models, HE16C, HE16L and HE16N
(representative of the models of the respective types-L, -M and H) and
performed a resolution study by progressively increasing the numerical
resolution. Our standard models have a working resolution $N_r\times
N_\theta=1000 \times 180$ zones. The standard computational grid is
uniform in both $\log{r}$- and $\theta$-coordinates. In addition to
the standard resolution, two higher resolutions have been considered.

First, we compute models HE16C-M, HE16L-M, and HE16N-M
(Tab.~\ref{tab:progenitor}), which have an intermediate resolution of
$N_r\times N_\theta=1500 \times 180$ zones. In these models the radial
grid is uniform in $\log{r}$ and the smallest radial grid spacing is
$\Delta r=7.5\times 10^{5}$\,cm, i.e., 3/4 times smaller than that of
our standard resolution cases. The polar grid possesses a uniform
region close to the symmetry axis ($0^{\circ}<\theta <40^{\circ}$),
where $\Delta \theta =1/3^{\circ}$, followed by a uniformly spaced
region in $\log{\theta}$ ($30^{\circ}<\theta <90^{\circ}$). The reason
to consider two different regions in the $\theta$-spacing is that all
our jet models develop cocoons whose angular extension is $\theta<
30^\circ$, i.e., all the dynamics develops in a wedge covered by a
finer mesh in our computational grid. Thus, an increased resolution in
the abovementioned wedge surrounding the polar axis yields an
effective increase of the numerical resolution in the whole
computational grid.
  
  Even higher resolution models (HE16C-H, HE16L-H, and HE16N-H) have
  also been computed. In this case, the grid consists of $N_r=2000$
  zones uniformly spaced in $\log{r}$, with a minimum radial grid
  spacing $\Delta r=5\times 10^{5}$\,cm, i.e., one half of the width
  of smallest radial spacing of the standard resolution case.  We take
  in this case $N_\theta=180$ zones, also split in two regions: a
  uniform grid in the interval $0^{\circ}<\theta <30^{\circ}$ ($\Delta
  \theta =0.25^{\circ}$), followed by a uniformly spaced region in
  $\log{\theta}$ ($30^{\circ}<\theta <90^{\circ}$).

  Going to even larger resolutions in the $\theta$-direction increases
  the total computational time up to prohibitive limits due to the
  increased number of time steps associated to the fulfillment of the
  Courant condition.

\begin{figure}
\begin{center}
\includegraphics[scale=0.6]{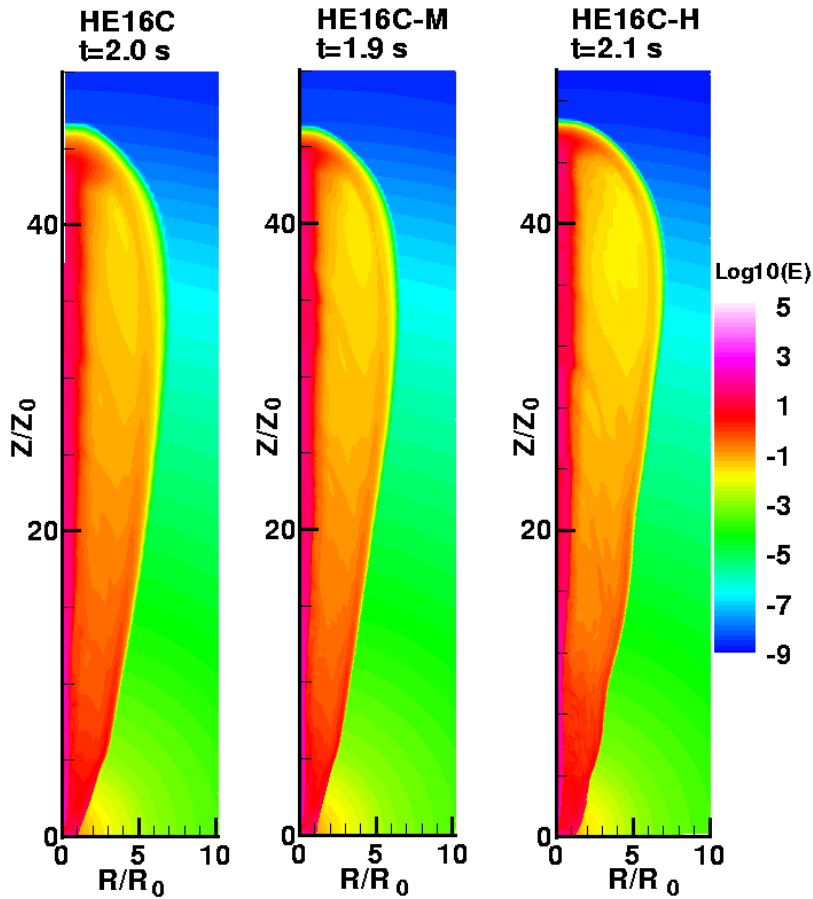}
\caption{Contours of the energy density for the same model run at three
  different resolutions, corresponding to models HE16C, HE16C-M, and
  HE16C-H. The jet in all cases is about to break the surface of the
  star ($R_* = 4.75\times 10^{10}$\,cm). Since the jet propagates at
  slightly different speed depending on the resolution, the three
  snapshots correspond to slightly different evolutionary times. Both, the vertical and the horizontal axis are scaled by $R_0=Z_0=10^9$\,cm.
  \label{fig:pressure_reso}}
\end{center}
\end{figure}

  We note that the jet dynamics is rather independent of the
  resolution, and also the gross morphological features are converged
  at the standard resolution, though, of course, finer details show up
  both in the cocoon and in the beam (see
  Fig.~\ref{fig:pressure_reso}). Therefore, when we take radial
  averages to compute the angular energy distribution of our models
  (Eq.~\ref{eq:dEdOmega}), the differences are relatively small (see
  below), and our standard resolution models can be considered to be
  sufficiently resolved to account for such global energetic
  properties. The time-scales to cross the progenitor and/or to reach
  the outer computational boundary (at $r=10^{11}$\,cm) are slightly
  different from each other. 

\begin{figure}
\begin{center}
\rotatebox{270}{\includegraphics[scale=0.35]{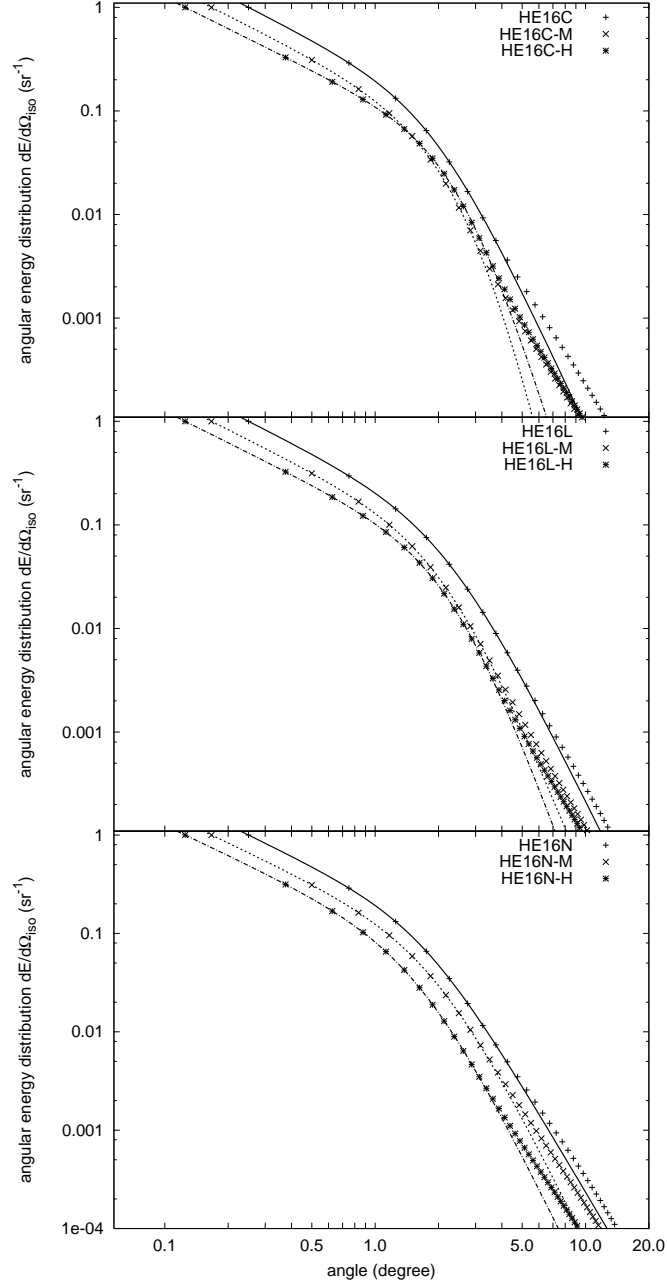}}
\caption{Comparison of the angular energy distributions of
prototype models (HE16C (top),
  HE16L (middle), and HE16L (bottom)) run with different
  computational resolutions.  The solid, dashed, and dotted lines
  show the fitting functions of Eq.~\ref{eq:fitting}
  corresponding to models computed with the standard, the
    middle, and the highest resolution, respectively.
    \label{fig:energy_distri_reso}}
\end{center}
\end{figure}

  Figure~\ref{fig:energy_distri_reso} shows the isotropic equivalent
  angular energy distributions of models computed with three different
  resolutions. In all cases, a SBP function (Eq.~\ref{eq:fitting})
  fits properly the data up to $\theta < 3.4^\circ$.  At larger
  latitude, the contribution of the expanding cocoon component
  dominates.  Since each distribution is normalized independently to
  its absolute maximum, the distributions corresponding to different
  resolutions do not overlap, but they show the same shape.  The
  fitting parameters for the higher resolution cases are also listed
  in Tab.~\ref{tab:progenitor}. There, we can see that the same
  correlations found in models with the standard resolution are
  reproduced at higher resolutions, namely, the correlation between
  $\alpha_{\rm h}$ or \ahiso with the progenitor mass
  (Sect.~\ref{sec:dEdOmega}). However, at higher resolution, we could
  guess additional correlations which are not obvious in models run
  with the standard resolution. Particularly, the parameters
  $\theta_0$ and $\theta_{0,\rm iso}$ become smaller as the progenitor
  mass increases. This trend is however, an artifact of the models
  chosen as prototypes of each mass type. They are such, that the
  radius of the progenitor roughly grows with the mass. As pointed out
  by \cite{Aloy00}, progenitors with larger radii provide a larger
  inertial confinement which prevents the lateral expansion of the
  jet. Hence, for the models chosen in this resolution study, a larger
  progenitor mass yields better collimated jets. Taking the whole
  sample of models, but run at higher resolution, the correlations
  between $\theta_0$ or $\theta_{0,\rm iso}$ would not exist.

\end{appendix}

\clearpage

%\begin{landscape}
\begin{table}
%\vspace{-2.3cm}
\begin{center}
  \caption{Properties of the pre-supernova models taken from
    \citet{Woosley06} for our study.  Columns 1-4, list the model
    name, the mass loss rate (in the same units as
    \citealp{Woosley06}), the total mass, and the radius at
    pre-supernova stage.  Columns 5-9, display the best parameters of
    the SBP (Eq.~\ref{eq:fitting}) used to fit the angular
    energy distribution per solid angle
    (Eq.~\ref{eq:dEdOmega}). Finally, columns 10-14, show the best fit
    parameters for the angular distribution of equivalent isotropic
    energy per solid angle.  The last six rows correspond to
      some prototype models which have been run with higher numerical
      resolution. Models whose name ends with "-M" and
      "-H" have been run with numerical resolutions of $1500\times180$
      and $2000\times180$ grid points, respectively (see Appendix).
\label{tab:progenitor}}
\begin{tabular}{cccc|ccccc|ccccc}
  \tableline\tableline
  model & mass & total           & Radius& $\alpha_l$  &
 $\alpha_h$& $A/10^{53}$ &
 $\theta_{0}$ & $n$ & ${\alpha_l}_{iso}$ & ${\alpha_h}_{iso}$& $A_{iso}/10^{57}$ &${\theta_{0}}_{iso}$ &$n_{iso}$\\ 
        &   loss        & mass /$M_\odot$ & /$10^{10}$ cm & /$10^{-2}$& & ergs& & & & & ergs& & \\ \tableline
HE16A & 0    & 15.70 & 3.86 & -1.22 & -3.27 & 8.65 & 2.37 & -0.686 & -1.01 & -4.24 & 2.45 & 2.35 & -0.696 \\
HE16B & 1.0  &  5.10 & 4.91 & -1.72 & -3.68 & 6.02 & 2.12 & -0.632 & -1.02 & -4.45 & 2.14 & 2.02 & -0.694 \\
HE16C & 1.0  &  5.15 & 4.75 & -2.99 & -3.58 & 6.61 & 2.03 & -0.706 & -1.03 & -4.38 & 2.40 & 1.95 & -0.765 \\
HE16D & 0.3  &  9.53 & 4.42 & -1.22 & -3.05 & 8.84 & 2.21 & -0.746 & -1.01 & -3.99 & 2.72 & 2.18 & -0.768 \\
HE16E & 0.1  & 12.86 & 5.71 & -2.60 & -2.73 & 9.64 & 1.84 & -0.896 & -1.03 & -3.64 & 3.64&1.80 & -0.942 \\
HE16F & 0.03 & 14.80 & 3.52 & -1.33 & -3.15 & 9.28 & 2.41 & -0.726 & -1.01 & -4.09 & 2.63&2.37 & -0.746 \\
HE16G & 0.01 & 15.56 & 3.31 & -1.27 & -3.26 & 8.55 & 2.44 & -0.719 & -1.01 & -4.30 & 2.24&2.47 & -0.705 \\
HE16H & 0    & 15.68 & 3.31 & -1.71 & -3.41 & 8.96 & 2.34 & -0.690 & -1.02 & -4.37 & 2.58&2.32 & -0.701 \\
HE16I & 0    & 15.88 & 4.54 & -1.21 & -3.17 & 8.12 & 2.33 & -0.714 & -1.01 & -4.14 & 2.32&2.32 & -0.722 \\ 
HE16J & 1.0  &  5.13 & 4.81 & -3.64 & -3.47 & 7.23 & 1.89 & -0.764 & -1.04 & -4.29 & 2.77&1.82 & -0.824 \\
HE16K & 1.0  &  5.16 & 4.81 & -3.49 & -3.57 & 6.67 & 2.02 & -0.741 & -1.04 & -4.38 & 2.39&1.95 & -0.798 \\
HE16L & 0.3  &  9.58 & 4.18 & -1.01 & -3.19 & 8.51 & 2.27 & -0.707 & -1.01 & -4.15 & 2.53&2.24 & -0.721 \\
HE16M & 0.1  & 13.04 & 6.29 & -2.29 & -2.80 & 8.89 & 1.93 & -0.851  & -1.03 & -3.69 & 3.24&1.88 & -0.903 \\
HE16N & 0.03 & 14.95 & 6.17 & -2.91 & -2.73 & 9.92 & 1.79 & -0.914  & -1.03 & -3.64 & 3.86&1.75 & -0.965 \\
HE16O & 0.01 & 15.62 & 5.96 & -1.88 & -2.74 & 9.11 & 1.99 & -0.855  & -1.02 & -3.67 & 3.16&1.94 & -0.891 \\
HE16P & 0    & 15.88 & 4.63 & -1.40 & -2.99 & 9.78 & 2.18 & -0.762 & -1.01 & -3.93 & 3.05&2.15 & -0.781 \\
16TB  & 0.1  & 15.29 & 4.45 & -1.42 & -3.14 & 8.38 & 2.19 & -0.733 & -1.01 & -4.11 & 2.55&2.18 & -0.741 \\
16TC  & 0.1  & 15.23 & 4.87 & -1.46 & -3.14 & 7.51 & 2.30 & -0.735 & -1.02 & -4.08 & 2.22&2.27  & -0.754 \\
16OC  & 0.1  & 14.26 & 4.42 & -1.62 & -2.95 & 9.05 & 2.09 & -0.798 & -1.02 & -3.94 & 2.85&2.09 & -0.798 \\ \tableline
HE16C-M & 1.0  &  5.15 & 4.75 & -0.72 & -5.77 & 6.71 & 2.70 & -0.486 &-1.01  & -9.81 & 0.310&3.43 & -0.241 \\
HE16L-M & 0.3  &  9.58 & 4.18 & -1.16 & -4.45 & 11.8 & 2.40 & -0.572 &-1.01  & -5.04 & 2.97&2.29 & -0.569 \\
HE16N-M & 0.03 & 14.95 & 6.17 & -1.27 & -3.33 & 17.9 & 1.65 & -0.685 &-1.01  & -4.41 & 3.64&2.06 & -0.645 \\ \tableline
HE16C-H & 1.0  &  5.15 & 4.75 & -1.99 & -6.18 & 4.40 & 2.64 & -0.377 &-1.00  & -7.76 & 1.08&2.98 & -0.397 \\
HE16L-H & 0.3  &  9.58 & 4.18 & -1.04 & -3.90 & 11.0 & 2.22 & -0.600 &-1.01  & -6.36 & 2.04&2.74 & -0.446 \\
HE16N-H & 0.03 & 14.95 & 6.17 & -1.18 & -3.22 & 12.4 & 1.96 & -0.704 &-1.01  & -4.62 & 5.74&1.79 & -0.596 \\ \tableline
\end{tabular}
\end{center}
\end{table}
%\end{landscape}

\end{document}